\documentclass[prd,twocolumn,eqsecnum,amsfonts,amssymb]{revtex4}

\usepackage{graphicx}

\usepackage{bm}

\newcommand{\beq}{\begin{equation}}
\newcommand{\eeq}{\end{equation}}
\newcommand{\beqs}{\begin{eqnarray}}
\newcommand{\eeqs}{\end{eqnarray}}
\newcommand{\lsim}{\mathrel{\raisebox{-
.6ex}{$\stackrel{\textstyle<}{\sim}$}}}

\begin{document}

\title{Search for an Ultraviolet Zero in the Seven-Loop Beta Function of the 
$\lambda \phi^4_4$ Theory}

\author{Robert Shrock}

\affiliation{C. N. Yang Institute for Theoretical Physics and \\
  Department of Physics and Astronomy \\
Stony Brook University, Stony Brook, NY 11794 }

\begin{abstract}

We investigate whether the seven-loop beta function of the $\lambda
\phi^4_4$ theory exhibits evidence for an ultraviolet zero.  In
addition to a direct analysis of the beta function, we calculate and
study Pad\'e approximants and discuss effects of scheme
transformations on the results. Confirming and extending our earlier
studies of the five-loop and six-loop beta functions, we find that in
the range of $\lambda$ where the perturbative calculation of the
seven-loop beta function is reliable, the theory does not exhibit
evidence for an ultraviolet zero.

\end{abstract}

%\pacs{11.10.-z,11.10.Hi}

\maketitle

% =======================================================================

\section{Introduction}
\label{intro_section}

In this paper we consider the renormalization-group (RG) behavior of
the $\lambda \phi^4$ field theory in $d=4$ spacetime dimensions, where
$\phi$ is a real scalar field.  This theory, commonly denoted
$\phi^4_4$, is described by the Lagrangian
\beq
{\cal L} = \frac{1}{2}(\partial_\nu \phi)(\partial^\nu \phi) 
- \frac{m^2}{2} \phi^2 - \frac{\lambda}{4!} \, \phi^4 \ . 
\label{lagrangian}
\eeq
The Lagrangian (\ref{lagrangian}) is invariant under the global
discrete ${\mathbb Z}_2$ symmetry $\phi \to -\phi$.  Quantum loop
corrections lead to a dependence of the physical quartic coupling
$\lambda = \lambda(\mu)$ on the Euclidean energy/momentum scale $\mu$
at which this coupling is measured.  The dependence of $\lambda(\mu)$
on $\mu$ is described by the RG beta function of the theory,
$\beta_\lambda = d\lambda/dt$, or equivalently,
$\beta_a = da/dt$, where $dt=d\ln\mu$ \cite{rg} and
\beq
a \equiv \frac{\lambda}{(4\pi)^2} \ . 
\label{a} 
\eeq
(The argument $\mu$ will often be suppressed in the notation.)  Since
we will investigate the properties of the theory for large $\mu$ in
the ultraviolet (UV), the value of $m^2$ will not play an important
role in our analysis.  For technical convenience, we assume that $m^2$ is
positive. At a reference scale $\mu_0$, the quartic coupling
$\lambda(\mu_0)$ is taken to be positive for the stability of the
theory.  The one-loop term in this beta function has a positive
coefficient, so that for small $\lambda$, $\beta_\lambda > 0$ and
hence as $\mu \to 0$, the coupling $\lambda(\mu) \to 0$, i.e., the
theory is infrared (IR)-free.  This perturbative result is in agreement
with nonperturbative approaches \cite{nonpert}; some reviews include
\cite{zjbook,kbook}.

The beta function $\beta_a$ has the series expansion
\beq
\beta_a = a\sum_{\ell=1}^\infty b_\ell \, a ^\ell \ .
\label{beta}
\eeq
The $n$-loop ($n\ell$) beta
function, denoted $\beta_{a,n\ell}$, is given by
Eq. (\ref{beta}) with the upper limit of the loop summation
index $\ell=n$ instead of $\ell=\infty$. 
The one-loop and two-loop terms in $\beta_a$ are independent of
the scheme used for regularization and renormalization, while terms of
loop order $\ell \ge 3$ are scheme-dependent \cite{bgz74,gross75}.
For the O($N$) $\lambda |\vec \phi|^4$ theory with an $N$-component
field, $\vec\phi = (\phi_1,...,\phi_N)$, the coefficients $b_1$,
$b_2$, and $b_3$ were calculated in \cite{bgz74}.  Higher-loop
coefficients $b_\ell$ with $\ell \ge 3$ have been computed using the
$\overline{\rm MS}$ minimal subtraction scheme \cite{hooft,msbar}. A
calculation of $b_5$ and discussion of earlier computations of $b_4$
and $b_5$ (e.g., \cite{vkt,cglt,glt}) was given in \cite{b345,kbook}.
The coefficient $b_6$ was calculated for $N=1$ in \cite{kp1} and for
general $N$ in \cite{kp}.  Most recently, the seven-loop coefficient
$b_7$ was calculated in \cite{schnetz}.  In analyzing the series
expansion (\ref{beta}), one recalls that it is an asymptotic expansion
and the large-order behavior has been the subject of extensive study
\cite{asymptotic}, including \cite{dunne} and references
therein. 

An interesting question is whether, for the region of $\lambda$ where
a perturbative calculation of $\beta_\lambda$ is reliable, this beta
function exhibits evidence for a zero at some (positive) value of the
quartic coupling. This would be an ultraviolet fixed point (UVFP) of
the renormalization group, i.e., as $\mu \to \infty$, $\lambda(\mu)$
would approach this value (from below).  In previous work we have
investigated this question up to the five-loop order for the O($N$)
$\lambda |\vec \phi|^4$ theory in \cite{lam} and up to the six-loop
order for the real $\lambda \phi^4$ theory in \cite{lam2} and the
O($N$) $\lambda |\vec \phi|^4$ theory in \cite{lam3}, finding evidence
against such a UVFP.  In the present paper, using the results of
\cite{schnetz}, we extend our analysis to the seven-loop level. Our
analysis in \cite{lam3} covered a large range of specific $N$ values
and also included an argument for the absence of a UV zero in the
(rescaled) $n$-loop beta function at large $N$ (see Eqs. (3.12)-(3.13) 
in \cite{lam3}).  Thus, it will suffice to focus on the $N=1$
theory here.

In view of this previous evidence against a UV zero in
$\beta_\lambda$ and associated UVFP in the O($N$)
$\lambda |\vec\phi|^4$ theory, it is worthwhile to mention one case where
an IR-free quantum field theory is known to have a UVFP, namely, the 
nonlinear O($N$) $\sigma$ model in $d=2+\epsilon$ 
spacetime dimensions. In this theory, an exact solution was obtained in the 
limit $N \to \infty$ with $\lambda(\mu) N = x(\mu)$ a fixed function of $\mu$
and yielded the beta function 
\beq
\beta_x = \frac{dx}{dt} = \epsilon x \Big ( 1 - \frac{x}{x_{_{UV}}} \Big ) 
\label{betax}
\eeq
for small $\epsilon$, where $x_{_{UV}}=2\pi\epsilon$ is a UV fixed point of the
renormalization group \cite{nlsm}.  Since the leading term in $\beta_x$ is
positive for $\epsilon > 0$, this theory is IR-free. 
Thus, in this nonlinear O($N$) $\sigma$
model in $d=2+\epsilon$ dimensions, the coupling $x(\mu)$ flows (monotonically)
from $x=0$ at $\mu=0$ to $x=x_{_{UV}}$ as $\mu \to \infty$.  
Note that by making $\epsilon \ll 1$ one can arrange that the UVFP at
$x_{_{UV}}=2\pi\epsilon$ occurs at an
arbitrarily small value of the scaled coupling $x$.

This paper is organized as follows.  In Section \ref{betafunction_section} we
  review some relevant background.  In Section \ref{zeros_section} we
  present the results of our analysis of the seven-loop beta function.
  Section \ref{pade_section} contains a further analysis of this question of
  a UV zero using Pad\'e approximants, while Section \ref{scheme_section}
  discusses effects of scheme transformations.  Our conclusions are
  given in Section \ref{conclusion_section}.

% ========================================================================

\section{Beta Function}
\label{betafunction_section}

The $n$-loop truncation of (\ref{beta}), denoted 
$\beta_{a,n\ell}$, is a polynomial in
$a$ of degree $n+1$ having an overall factor of $a^2$.  We may 
extract this factor and define a reduced beta function 
\beqs
\beta_{a,r} &=& \frac{\beta_a}{\beta_{a,1\ell}} = \frac{\beta_a}{b_1 a^2}
  \cr\cr
  &=&
 1 + \frac{1}{b_1} \, \sum_{\ell=2}^\infty b_\ell a^{\ell-1} \ . 
\label{betared}
\eeqs
The $n$-loop truncation of $\beta_{a,r}$, denoted 
$\beta_{a,r,n\ell} \equiv R_n$, is defined by taking the upper limit of the
sum in (\ref{betared}) to be $\ell=n$ rather than $\ell=\infty$. .

The first two coefficients in the beta function of this theory are
$b_1=3$ and $b_2=-17/3$ \cite{bgz74}. The coefficients $b_\ell$ with
$3 \le \ell \le 7$ and the resultant higher-loop beta function
discussed below, are calculated in the $\overline{\rm MS}$ scheme.
The coefficients up to the five-loop level are
\cite{vkt,bgz74,b345,kbook}
\beq
b_3=\frac{145}{8} + 12\zeta_3 = 32.5497 \ , 
\label{b3}
\eeq
\beqs
b_4 &=& -\frac{3499}{48} - 78\zeta_3 + 18\zeta_4 - 120\zeta_5 \cr\cr
&=& -271.606 \ ,
\label{b4}
\eeqs
and
\beqs
b_5 & = & \frac{764621}{2304} + \frac{7965}{16}\zeta_3 - \frac{1189}{8}\zeta_4
+987\zeta_5 + 45\zeta_3^2 \cr\cr
& - & \frac{675}{2}\zeta_6+1323\zeta_7 \cr\cr
&=& 2848.57 \ , 
\label{b5} 
\eeqs
where the floating-point values are given to the indicated accuracy and  
\beq
\zeta_s = \sum_{n=1}^\infty \frac{1}{n^s}
\label{zeta}
\eeq
is the Riemann zeta function.
If $s=2r$ is even, then $\zeta_s$ can be expressed as a rational number times 
$\pi^{2r}$, namely $\zeta_{2r}=(-1)^{r+1}B_{2r}(2\pi)^{2r}/[2(2r)!]$, where 
$B_n$ are the Bernoulli numbers;
however, we leave these $\zeta_{2r}$ in their generic form here and below.  
% Numerically, $b_3=32.54968$, $b_4=-271.60578$, $b_5=2848.568256$.
The six-loop coefficient is \cite{kp1,kp}
\begin{widetext}
\beqs
b_6 & = & -\frac{18841427}{11520} - \frac{779603}{240}\zeta_3 
+ \frac{16989}{16}\zeta_4 -\frac{63723}{10}\zeta_5 -\frac{8678}{5}\zeta_3^2
+ \frac{6691}{2}\zeta_6 + 162\zeta_3\zeta_4 - \frac{63627}{5}\zeta_7 \cr\cr
& - & 4704\zeta_3\zeta_5 + \frac{264543}{25}\zeta_8 -
\frac{51984}{25}\zeta_{3,5}  - 768\zeta_3^3 - \frac{46112}{3}\zeta_9
\cr\cr
&=& -34776.13\ ,  
\eeqs
\end{widetext}
%
% Numerically, $b_6=-34776.13128$. 
where \cite{bbv}
\beq
\zeta_{3,5}=\sum_{m > n \ge 1} \frac{1}{n^3 m^5} \ .  
\label{zeta35}
\eeq
The seven-loop coefficient is considerably more complicated than $b_6$, and
we refer the reader to \cite{schnetz} for the analytic expression.  The
numerical value is
\beq
b_7 = 474651.0 \ .
\label{b7value}
\eeq
Thus, in summary, the seven-loop beta function of the $\lambda \phi^4$
theory (calculated in the $\overline{\rm MS}$ scheme), is 
\beqs
\beta_{a,7\ell} & = & a^2\Big ( 3 - \frac{17}{3}a + 32.5497a^2 -271.606a^3 
\cr\cr
&+& 2848.57 a^4 -34776.1 a^5 + 474651 a^6 \Big ) \ . \cr\cr
&&
\label{beta_6loop}
\eeqs
%
% ========================================================================

\section{Zeros of the $n$-Loop Beta Function up to Loop Order $n=7$ }
\label{zeros_section}

In this section we investigate a possible UV zero, denoted
$a_{_{UV,n\ell}}$, of the $n$-loop beta function, $\beta_{a,n\ell}$. The
double zero of $\beta_{a,n\ell}$ at $a=0$ is always present (independent of 
$n$); this is an infrared zero and hence will not be of interest here. 

A necessary condition for there to be robust evidence for a UV zero in
the beta function of an IR-free theory is that the values calculated
at successive loop orders should be close to each other. Although the
two-loop beta function $\beta_{a,2\ell}$ does have a UV zero, at
$a_{_{UV,2\ell}}= 9/17 = 0.52941$, we found that the three-loop beta
function $\beta_{a,3\ell}$ has no UV zero and, while a UV zero is
present in $\beta_{a,4\ell}$, it occurs at a considerably smaller
value, namely $a_{_{UV,4\ell}}=0.23332$.  At the five-loop level,
$\beta_{a,5\ell}$ has no UV zero, while at the six-loop level,
although $\beta_{a,6\ell}$ has a UV zero, it occurs at a still smaller
value, $a_{_{UV,6\ell}}=0.16041$ \cite{lam,lam2}. Thus, the results of
this analysis show that the necessary condition that the beta function
calculated to successively higher loop order should exhibit values of
$a_{_{UV,n\ell}}$ that are close to each other is not satisfied by
this theory. At seven-loop order, using $\beta_{a,7\ell}$ from
\cite{schnetz}, we find that this function has no physical UV zero.
Instead, the zeros are comprised of three complex-conjugate pairs,
$-0.102135 \pm 0.079848i$, $0.0142348 \pm 0.136854i$, and $0.124533 \pm
0.0659940i$. Summarizing,
\beqs
&& a_{_{UV,2\ell}}=0.52941, \quad a_{_{UV,4\ell}}=0.23332, \quad
a_{_{UV,6\ell}}=0.16041 \cr\cr
&& {\rm no} \ a_{_{UV,n\ell}} \ {\rm for} \ n=3, \ 5, \ 7.
\label{auvsummary}
\eeqs
The calculations up to seven loops show a pattern, namely
that for even $n=2, \ 4, \ 6$, $\beta_{a,n\ell}$ has a zero,
$a_{_{UV,n\ell}}$, but the values for different $n$
are not close to each other, while
for odd $n=1, \ 3, \ 5, \ 7$, $\beta_{a,n\ell}$ has no UV zero.

\begin{figure}
  \begin{center}
    \includegraphics[height=7cm,width=7cm]{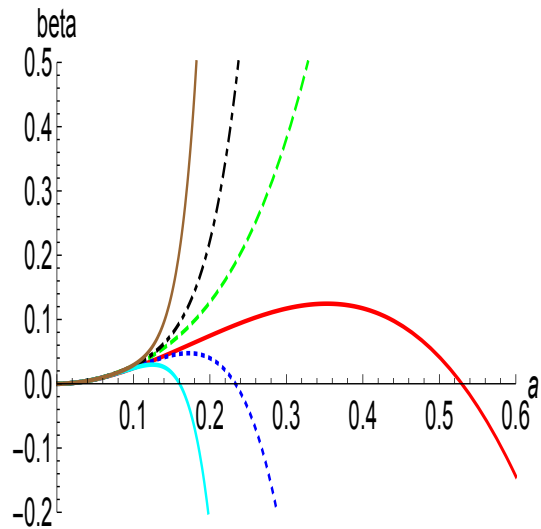}
  \end{center}
\caption{\footnotesize{ Plot of the $n$-loop $\beta$ function
    $\beta_{a,n\ell}$ as a function of $a$ for (i) $n=2$ (red, solid),
    (ii) $n=3$ (green, dashed), (iii) $n=4$ (blue, dotted), (iv) $n=5$
    (black, dot-dashed), (v) $n=6$ (cyan, solid), and (vi) $n=7$
    (brown, solid).  At $a=0.16$, going from bottom to top, the curves
    are for $n=6$, $n=4$, $n=2$, $n=3$, $n=5$, $n=7$.}}
\label{beta_lam_Neq1}
\end{figure}

\begin{figure}
  \begin{center}
    \includegraphics[height=7cm,width=7cm]{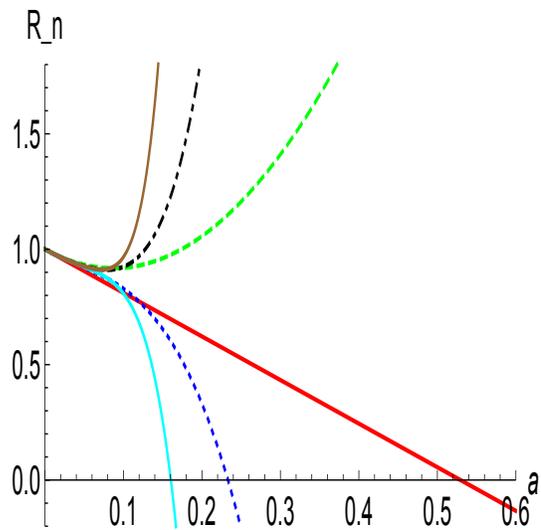}
  \end{center}                             
\caption{\footnotesize{Plot of the ratio $R_n$ of the $n$-loop beta
    function $\beta_{a,n\ell}$ divided by $\beta_{a,1\ell}$, as a
    function of $a$ for (i) $n=2$ (red, solid), (ii) $n=3$ (green,
    dashed), (iii) $n=4$ (blue, dotted), (iv) $n=5$ (black,
    dot-dashed), (v) $n=6$ (cyan, solid), and (vi) $n=7$ (brown,
    solid).  At $a=0.16$, going from bottom to top, the curves are for
    $n=6$, $n=4$, $n=2$, $n=3$, $n=5$, and $n=7$.}}
\label{betar_lam_Neq1}
\end{figure}

In Fig. \ref{beta_lam_Neq1} we plot the $n$-loop beta functions for $2
\le n \le 7$ loops. Another way to show this information is via the
$n$-loop reduced beta function, $\beta_{a,r,n\ell} = R_n$.  We plot
$R_n$ in Fig. \ref{betar_lam_Neq1} for $2 \le n \le 7$.  The results
discussed above are evident in these figures.  First, one may inquire
how large is the interval in $a$ over which the calculations of
$\beta_{a,n\ell}$ to the respective $n$-loop orders are in mutual
agreement. As one can see from Figs. \ref{beta_lam_Neq1} and
\ref{betar_lam_Neq1}, the $n$-loop beta functions $\beta_{a,n\ell}$
with $2 \le n \le 7$ only agree with each other well over the small
interval of couplings $0 \le a \lsim 0.05$.  As shown in
Fig. \ref{beta_lam_Neq1}, the $\beta_{a,n\ell}$ with even $n=2, \ 4,
\ 6$ reach maxima and then decrease, crossing the (positive) real axis
at different values listed in Eq. (\ref{auvsummary}), while the
$\beta_{a,n\ell}$ with odd $n$ increase monotonically with $a$.  This
seven-loop analysis confirms and extends our conclusions in
\cite{lam2,lam3} at the six-loop level that the zero in the two-loop
beta function of the $\lambda\phi^4$ theory occurs at too large a
value of $a$ for the perturbative calculation to be reliable.

%

% ======================================================================

\section{Analysis With Pad\'e Approximants}
\label{pade_section}

One can gain further insight into the behavior of the beta function by
the use of Pad\'e approximants (PAs).  We carried out
this analysis up to the six-loop level in \cite{lam2,lam3}, finding no
indication of a physical UV zero, and here we extend it to the
seven-loop level.  Since the double zero in $\beta_{a,n\ell}$ at $a=0$
is not relevant to the question of a UV zero, we use the reduced beta
function $\beta_{a,r,n\ell}$ for this Pad\'e analysis.  The $[p,q]$
Pad\'e approximant to $\beta_{a,r,n\ell}$ is the rational function
\cite{pade1}
\beq
[p,q]_{\beta_{a,r,n\ell}} =
\frac{1+\sum_{j=1}^p \, r_j a^j}{1+\sum_{k=1}^q s_k \, a^k}
\label{pqx}
\eeq
with $p+q=n-1$, 
where the coefficients $r_j$ and $s_j$ are independent of $a$.
At seven-loop order, we can calculate the Pad\'e approximants
$[p,q]_{\beta_{a,r,7\ell}}$ with $[p,q]$ taking on the values
[6,0], [5,1], [4,2], [3,3], [2,4], [1,5], and [0,6].  Since the loop order
is understood, we write $[p,q]_{\beta_{a,r,7\ell}} \equiv [p,q]$ for brevity
of notation. The PA [6,0] is equivalent to $\beta_{a,r,7\ell}$ itself,
which we have already analyzed, and the PA [0,6] has no zeros, so we focus
here on the remaining five Pad\'e approximants. 

We list our results for these Pad\'e approximants to $\beta_{a,r,7\ell}$ below:
\begin{widetext}
\beq
[5,1] = \frac{1+11.760a-14.931a^2+57.552a^3-286.17a^4+1367.8a^5}
{1+13.649a} \ ,
\label{bpade51}
\eeq
\beq
    [4,2] = \frac{1+20.541a+75.687a^2-49.670a^3+81.973a^4}
    {1+22.430a+107.21a^2} \ ,
\label{bpade42}
\eeq
\beq
    [3,3] = \frac{1+25.073a+152.81a^2+155.99a^3}
    {1+26.962a+192.89a^2 +318.33a^3} \ ,
\label{bpade33}
\eeq
\beq
    [2,4] = \frac{1+22.314a+103.55a^2}
    {1+24.203a+138.42a^2+89.390a^3-91.252a^4} \ ,
\label{bpade24}
\eeq
\beq
    [1,5] = \frac{1+14.023a}
    {1+15.912a+19.205a^2 -45.828a^3 +196.10a^4 -910.03a^5} \ . 
\label{bpade15}
\eeq
\end{widetext}
We recall some necessary requirements for a zero of a $[p,q]$ Pad\'e
approximant to be physically relevant. These include the requirement
that this zero should occur on the positive real axis in the complex
$a$ plane and the requirement that this zero of the PA should be
closer to the origin $a=0$ than any pole on the real positive
$a$-axis, since otherwise the pole would dominate the IR to UV flow
starting at the origin.  If a Pad\'e approximant were to exhibit such
a zero, then one would proceed to inquire how close it is to any of
the $a_{_{UV,n\ell}}$ in Eq. (\ref{auvsummary}).  However, we find
that none of these Pad\'e approximants (\ref{bpade51})-(\ref{bpade15})
has a zero on the positive real $a$ axis. Explicitly, the [5,1] PA has two
complex-conjugate pairs of zeros at $a=-0.12719 \pm 0.26046i$ and
$a=0.26922 \pm 0.20930i$, together with a real zero at
$a=-0.074837$. This real zero is part of a nearly coincident pole-zero pair,
with the pole of the [5,1] PA being located at $a=-0.073267$. The
appearance of a nearly coincident pole-zero pair close to a point
$a_0$ in a $[p,q]$ Pad\'e approximant is typically an indication that
the function that the PA is fitting has neither a pole nor a zero in
the local neighborhood of $a_0$, since as the locations of the nearly
coincident pole-zero pair approach each other, they simply divide out
in the ratio (\ref{pqx}).  Each of the Pad\'e approximants that we
calculate here has a pole-zero pair.  The [4,2] PA has zeros at the
complex-conjugate pair $a=0.42009 \pm 0.96575i$, together with the
real values $a=\{-0.16929, \ -0.064970\}$ and poles at $a=\{-0.14481,
\ -0.064414\}$. The [3,3] PA has zeros at $a=\{-0.78531, \ -0.13282,
-0.061458\}$, and poles at $a=\{-0.42342, \ -0.12140,
\ -0.061112\}$. The [2,4] PA has zeros at $a=\{-0.15193,
\ -0.063563\}$, and poles at $a=\{-0.69186, \ -0.13432, \ -0.063100,
\ 1.8689\}$. Finally, the [1,5] PA has a zero at $a=-0.071313$ and
poles at $a=\{-0.22780, \ -0.070185, \ 0.44160, \ 0.035937 \pm
0.39287i\}$.  Thus, our analysis with Pad\'e approximants of the
seven-loop beta function yields the same conclusion as our analysis of
the beta function itself, namely that there is no evidence for a
stable, reliably perturbatively calculable UV zero up to this
seven-loop level.

% ========================================================================

\section{Effects of Scheme Transformations} 
\label{scheme_section}

Since the terms in the beta function at loop order $n \ge 3$ are
scheme-dependent, it is necessary to assess the effect of scheme
transformations in an analysis of zeros of a higher-loop beta function.
A scheme transformation can be expressed as a mapping between $a$ and
a transformed coupling $a'$,
\beq
a = a' f(a') \ ,
\label{aap}
\eeq
where $f(a')$ is the scheme transformation function. Since this transformation
has no effect in the free theory, one has $f(0) = 1$.  We consider $f(a')$
functions that are analytic about $a=a'=0$ and hence can be expanded in the
form
\beq
f(a') = 1 + \sum_{s=1}^{s_{max}} k_s (a')^s \ , 
\label{faprime}
\eeq
where the $k_s$ are constants and $s_{max}$ may be finite or infinite. The
beta function in the transformed scheme, $\beta_{a'}=da'/d\ln\mu$, has
the expansion
\beq
\beta_{a'} = a'\sum_{\ell=1}^\infty b_\ell' (a')^\ell \ . 
\label{betaprime}
\eeq
In \cite{sch}, formulas were derived for the $b_\ell'$ in terms of
$b_\ell$ and the $k_s$. In addition to $b_1'=b_1$ and $b_2'=b_2$,
these are 
\beq
b_3' = b_3 + k_1b_2+(k_1^2-k_2)b_1 \ ,
\label{b3prime}
\eeq
\beq
b_4' = b_4 + 2k_1b_3+k_1^2b_2+(-2k_1^3+4k_1k_2-2k_3)b_1 \ ,
\label{b4prime}
\eeq
and so forth for higher $\ell$.  These results are applicable to the
study of both an IR zero in the beta function of an asymptotically
free theory and a possible UV zero in the beta function of an IR-free
theory. They were extensively applied to assess scheme dependence in
higher-loop studies of an IR fixed point in asymptotically free
non-Abelian gauge theories \cite{sch,sch23,tr,schl,gracey_simms}.

For the present $\lambda\phi^4$ theory, a study of scheme dependence
was carried out in \cite{lam}.  It was shown that even when one shifts
to a scheme different from the usual $\overline{\rm MS}$ scheme, the
beta function still does not satisfy a requisite condition for a physical UV
zero, namely that the value of this zero (in a given scheme) should
not change strongly when it is calculated to successive loop
orders. This result from \cite{lam} also holds in the same way in the
present seven-loop context.

% =======================================================================
\section{Conclusions} 
\label{conclusion_section}

In this paper we have investigated whether the real scalar field
theory with a $\lambda \phi^4$ interaction exhibits evidence of an
ultraviolet zero in the beta function. Using the seven-loop
coefficient $b_7$ from \cite{schnetz}, our present study extends our
previous six-loop study in \cite{lam2,lam3} to the seven-loop level.
Our work includes a study of the seven-loop beta function itself,
together with an analysis of Pad\'e approximants.  We conclude that,
for the range of couplings where the perturbative calculation of this
beta function may be reliable, it does not exhibit robust evidence for
an ultraviolet zero.

% =======================================================================

\begin{acknowledgments}

I would like to thank Oliver Schnetz for valuable discussions on
\cite{schnetz}.  This research was supported in part by the
U.S. National Science Foundation Grant NSF-PHY-22-15093.

\end{acknowledgments}

% ================================================================

\end{document}